\documentclass{ws-procs975x65}
\usepackage{graphicx}
\usepackage{amsmath}
\usepackage{amssymb}
\usepackage{enumerate}
\usepackage{epsfig}
\usepackage{verbatim}
\usepackage{array}
\makeatletter

\def\vereq#1#2{\lower3pt\vbox{\baselineskip0.5pt\lineskip0.5pt
\ialign{$\m@th#1\hfill##\hfil$\crcr#2\crcr\sim\crcr}}}

\makeatother
\newcommand{\lsim}{\lower.5ex\hbox{$\; \buildrel < \over\sim \;$}}
\newcommand{\gsim}{\lower.5ex\hbox{$\; \buildrel > \over\sim \;$}}
\newcommand{\postscript}[2]{\setlength{\epsfxsize}{#2\hsize}
   \centerline{\epsfbox{#1}}}

\newcommand{\lwig}{\mbox{\,\raisebox{.3ex}
    {$<$}$\!\!\!\!\!$\raisebox{-.9ex}{$\sim$}\,}}
\newcommand{\gwig}{\mbox{\,\raisebox{.3ex}
    {$>$}$\!\!\!\!\!$\raisebox{-.9ex}{$\sim$}}\,}

\newif\ifarxiv

\arxivtrue
\begin{document}

\title{
\ifarxiv
\vspace{-3cm}
{\rm\normalsize\rightline{NUB-3248/Th-04}\rightline{DESY 04-031}\rightline{\lowercase{hep-ph/0403001}}}
\vskip 1cm 
\fi
Frontiers in Cosmic Rays\ifarxiv\footnote{
  \uppercase{R}apporteur review of \uppercase{HEA}2 --
  10th \uppercase{M}arcel \uppercase{G}rossmann \uppercase{M}eeting on 
  \uppercase{G}eneral \uppercase{R}elativity --}\fi }

\author{Luis A. Anchordoqui}
\address{Department of Physics,
Northeastern University, Boston, MA 02115, USA\\
E-mail: {\tt l.anchordoqui@neu.edu}}
\author{Charles D. Dermer}
\address{Code 7653, Naval Research Laboratory, 4555 Overlook Ave.\ SW,  
Washington, DC 20375-5352 USA\\
E-mail: {\tt dermer@gamma.nrl.navy.mil}}
\author{Andreas Ringwald}
\address{Deutsches Elektronen-Synchrotron DESY, D-22603 Hamburg, Germany\\
E-mail: {\tt andreas.ringwald@desy.de}}


\maketitle

\abstracts{This rapporteur review covers selected results presented in 
 the Parallel Session HEA2 (High Energy Astrophysics 2) of the {\it
10th Marcel Grossmann Meeting on General Relativity}, held in Rio de
Janeiro, Brazil, July 2003. The subtopics are: ultra high energy
cosmic ray anisotropies, the possible connection of these energetic
particles with powerful gamma ray bursts, and new exciting scenarios
with a strong neutrino-nucleon interaction in the atmosphere.}

\section{Introduction}

Since the early 60's several ground-based experiments have observed
extensive air showers, presumably triggered by ultra high energy
cosmic rays (UHECRs) interacting in the upper
atmosphere.\cite{Anchordoqui:2002hs} The highest primary energy
measured thus far is $E \sim 10^{20.5}$~eV,\cite{Bird:1994uy}
corresponding to a center-of-mass energy $\sqrt{s} = \sqrt{2 m_p E}
\sim 750$~TeV, where $m_p$ is the proton mass. The interest in the
origin of these particles is twofold: there is not only the
intellectual curiosity about unknown properties of powerful
astrophysical scenarios, but also the possibility to probe new physics
at energies beyond the reach of any foreseeable man-made experiments.

\begin{figure} [t]
\postscript{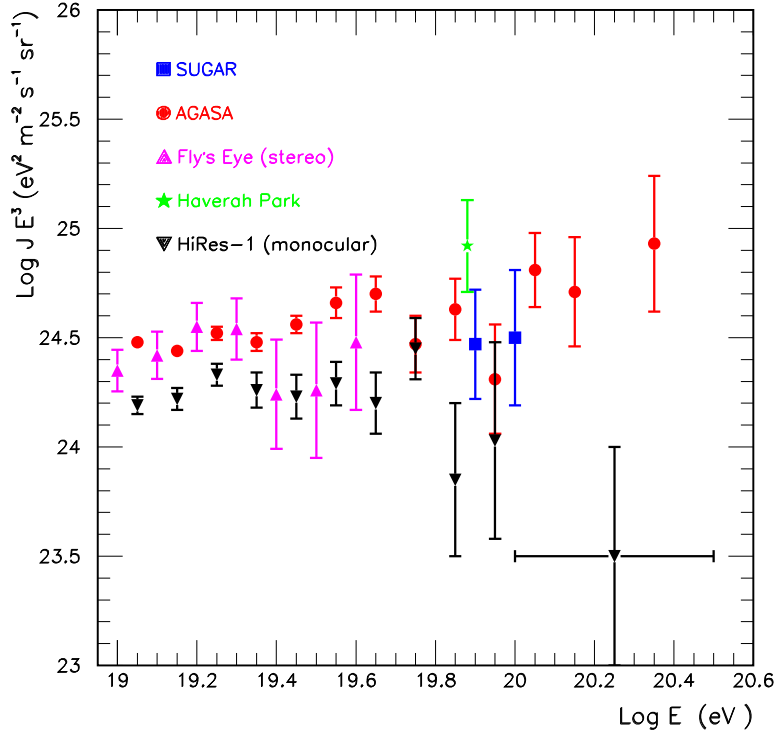}{0.80}
\caption[...]{Data on the upper end of the cosmic ray energy spectrum 
 with  their statistical error bars. (HiRes,\cite{Abu-Zayyad:2002sf}
AGASA,\cite{Takeda:2002at} Fly's Eye,\cite{Bird:wp} Haverah
Park,\cite{Ave:2001hq} and SUGAR.\cite{Anchordoqui:2003gm})
\label{azukar}}
\end{figure}

Theoretically, one expects the CR spectrum to fall off somewhat above
$10^{20}$~eV, because the particle's energy gets degraded through
interactions with the cosmic microwave background (CMB), a phenomenon
known as the Greisen-Zatsepin-Kuzmin (GZK)
cutoff.\cite{Greisen:1966jv} Unfortunately, as one can see in
Fig.~\ref{azukar}, the most
recent measurements by the HiRes\cite{Abu-Zayyad:2002sf} and
AGASA\cite{Takeda:2002at} experiments are apparently in conflict, if
only statistical errors are taken into account, and the source of the
difference remains unknown. However, if one takes the systematic
uncertainties in the energy measurements into account, one finds that
both data sets are mutually compatible on the $2\,\sigma$
level.\cite{DeMarco:2003ig} Attempts to explain the AGASA data with a
homogeneuos population of astrophysical sources that injects power-law
distributions of CRs give unacceptable $\chi^2$ (see, e.g.,
Refs.~[\cite{Fodor:2003bn,Fodor:2003ph,Wick:2003ex}]).  On the other
hand, an analysis\cite{Bahcall:2002wi} of the combined data reported
by the HiRes, the Fly's Eye, and the Yakutsk collaborations is
supportive of the existence of the GZK cutoff at the $>5\,\sigma$
($>3.7\,\sigma$, depending upon the extrapolated energy spectrum)
level.\footnote{This evidence disappears, however, if one assumes that
UHECRs are protons and excludes nearby ($\lwig\, 50$~Mpc) sources from
the otherwise homogeneous distribution.\cite{Kachelriess:2003yy} In
this case even the HiRes-1 data are incompatible with the GZK cutoff
on the $3\,\sigma$ level.\cite{Fodor:2003bn,Fodor:2003ph}} The
deviation from GZK depends on the set of data used as a basis for
power law extrapolation from lower energies. One caveat is a recent
claim\cite{Watson} that there may be technical problems with the
Yakutsk data collection. In view of the low statistics at the end of
the spectrum and the wide variety of uncertainties in these
experiments, perhaps the rational thing to do is to wait for more data,
conservatively arguing that the jury is still out.

In this Parallel Session we saw many thorough reviews covering all the
most interesting and timely topics in CR physics.  In this rapporteur
summary we cannot do justice to all the presentations.\footnote{A
scenario in which UHECRs are able to break the GZK barrier was
presented by She-Sheng Xue.\cite{Xue:2002hq}} Priority will be given
to two intriguing scenarios, which pose possible explanations of the
data.

\section{Anisotropies in UHECRs}
\label{john}

At the highest energies, the arrival directions of CRs are expected to
begin to reveal their origins. If the CR intensity were isotropic,
then one should expect a time-independent flux from each direction in
local detector coordinates, i.e., declination and hour angle.  In that
case, a shower detected with local coordinates could have arrived with
equal probability at any other time of a shower detection. For any
point of the celestial sphere, the expected shower density can be
estimated if the exposure in each direction can be obtained.  This
implies that celestial anisotropies can be easily discerned by
comparing the observed and expected event frequencies at each region.

For experiments with 100\,\% duty cycle, continuous operation in solar
time for several years leads to a uniform observation in right
ascension. Therefore, one of the conventional methods to search for
any global anisotropy is to apply the Linsley's\cite{Linsley:aniso}
harmonic analysis to the full sky cosmic ray distribution, i.e.,
determine the amplitude and phase of the $m^{\rm th}$ harmonic by
fitting the right ascension distribution of events to a sine wave with
period $2\pi/m.$

There is a remarkable agreement among several experiments favoring a
significant anisotropy (encoded in the first harmonic amplitude)
around $10^{18}$~eV from the general direction of the Galactic Plane
(GP). Specifically, the AGASA experiment has revealed a correlation
between the arrival direction of CRs (with energy $\sim 10^{18}$~eV)
and the GP at the $4\,\sigma$ level.\cite{Hayashida:1998qb} The GP
excess, which is roughly 4\% of the diffuse flux, is mostly
concentrated in the direction of the Cygnus region, with a second spot
towards the Galactic Center (GC).\cite{Teshimaicrc} Evidence at the
3.2$\,\sigma$ level for GP enhancement in a similar energy range has
also been reported by the Fly's Eye Collaboration.\cite{Bird:1998nu}
Interestingly, the full Fly's Eye data include a directional signal
from the Cygnus region which was somewhat lost in an unsuccessful
attempt to relate it to $\gamma$-ray emission from Cygnus
X-3.\cite{Cassiday:kw} Finally, the existence of a point-like excess
in the direction of the GC has been confirmed via independent analysis
of data collected with the SUGAR experiment.\cite{Bellido:2000tr}

For the ultra high energy~($>10^{19.6}$~eV) regime, all experiments to
date have reported no departure from isotropy in the first harmonic
amplitude.\cite{Edge:rr}\footnote{For the Fly's Eye data-sample the
first harmonic amplitude is computed using weighted showers, because
it has had a nonuniform exposure in sideral time. A shower's weight
depends on the hour of its sideral arrival time, and the 24 different
weights are such that every time bin has the same weighted number of
showers.}  This does not imply an isotropic distribution, but it
merely means that available data are too sparse to claim a
statistically significant measurement of anisotropy.  In other words,
there may exist anisotropies at a level too low to discern given
existing statistics.\cite{Evans:2001rv}

The right ascension harmonic analyses are completely blind to
intensity variations which depend only on declination.  Combining
anisotropy searches in right ascension over a range of declinations
could dilute the results, since significant but out of phase
``Rayleigh vectors'' from different declination bands can cancel each
other out.  Moreover, the analysis methods that consider distributions
in one celestial coordinate, while integrating away the second, have
proved to be potentially misleading.\cite{Wdowczyk:rb} An unambiguous
interpretation of anisotropy data requires two ingredients: {\it
exposure to the full celestial sphere and analysis in terms of both
celestial coordinates.}\cite{Sommers:2000us}

The first full sky search for large scale anisotropies in the
distribution of arrival directions of CRs with energy $> 10^{19.6}$~eV
was reported in this Parallel Session by John
Swain.\cite{Swain:2004na} Data from the SUGAR and AGASA experiments,
taken during a 10~yr period with nearly uniform exposure to the entire
sky, show no departures from either homogeneity nor isotropy on
angular scale greater than $10^\circ$.

\begin{figure} [t]
\vspace{-4ex}
\postscript{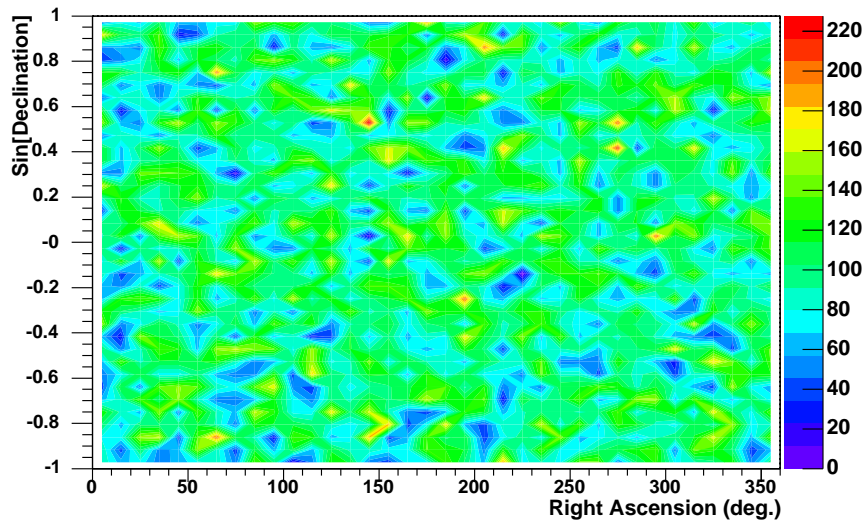}{0.90}
 \caption{UHECR intensity in arbitrary units (equatorial 
coordinates) as seen by the AGASA and SUGAR experiments.   
\label{intensity} }
\end{figure}

In this full-sky anisotropy search, the intensity distribution of the
set of $N = 99$ arrival directions
\begin{equation}
I({\bf{n}}) = \frac{1}{\mathcal N}\,\,\sum_{j = 1}^N  \frac{1}{\omega_j} \,\, 
\delta ({\bf n}, {\bf n}_j) \,\,, 
\label{I}
\end{equation}
was conviniently expanded in spherical harmonics ($Y_{\ell m}$)
\begin{equation}
I({\bf n})= \sum_{\ell = 0}^\infty \,\, \sum_{m = -\ell}^\ell\,\, a_{\ell m}\,  Y_{\ell m}({\bf n})\,\,,
\label{CP20}
\end{equation}
going into the multipole expansion out to $\ell =5$.  Here, $\omega_j$
is the relative exposure at arrival direction ${\bf n}_j$ and
${\mathcal N}$ is the sum of the weights $\omega_j^{-1}$.  The
coordinate independent total power spectrum of fluctuations,
\begin{equation}
C(\ell) = \frac{1}{(2 \ell +1)}\,\, \sum_{m=-\ell}^\ell a_{\ell m}^2\,\,,
\label{CP23}
\end{equation}
is consistent with that expected from a random distribution for all
(analyzed) multipoles, though there is a small ($2\sigma$) excess in
the data for $\ell =3$.\cite{Anchordoqui:2003bx} To give a visual
impression of the level of homogeneity and isotropy in existing data,
in Fig.~\ref{intensity} we show the intensity distribution as seen by
AGASA and SUGAR experiments.

\section{UHECRs from GRBs }

In this section, arguments for the origin of UHECRs from gamma ray
bursts (GRBs) are reviewed. This line of enquiry has led to a complete
model for CRs originating from supernovae (SNe) and GRBs in our Galaxy
and throughout the universe,\cite{Dermer:2000yd,Wick:2003ex} which is
summarized here.

The connection betweeen GRBs and UHECRs was first made on the basis of
an intriguing coincidence\cite{Vietri:1995hs,Waxman:1995vg} beween the
power required to sustain the measured flux of super-GZK ($\gsim
10^{20}$ eV) CRs against photohadronic energy losses and the local
time- and space-averaged hard X-rays/soft $\gamma$-ray luminosity of
GRBs. This luminosity density is estimated to be $\approx 10^{44}$
ergs Mpc$^{-3}$ yr$^{-1}$.  Thus GRBs have, in principle, sufficient
energy to power the UHECRs and post-GZK CRs. Moreover, many GRB
sources are found within the GZK radius, and CRs with energies $\gsim
10^{20}$ eV can be accelerated by the relativistic shocks formed in
GRB explosions.\cite{Vietri:2003te} The hypothesis of a GRB/UHECR
association points to a closer connection between SNe and CRs that
could provide a complete solution to the problem of CR origin.

\subsection{CRs from Supernovae}

Even though the controversy surrounding the origin of the UHECRs has
generated much interest, it should be noted that the much older
problem of the origin of the CRs is itself not solved. Cosmic rays
with energies from GeV/nucleon up to hundreds of TeV are widely
thought to be accelerated by supernova remnant (SNR) shocks. Yet the
prediction that SNRs should be luminous $\gamma$-ray sources and
display the characteristic 70 MeV $\pi^0$ decay emission feature from
hadronic interactions was not confirmed by the EGRET instrument on the
{\it Compton Observatory}. Nevertheless, there is statistical evidence
that SNRs are associated with unidentified $\gamma$-ray
sources.\cite{Torres:2002af} There is also clear evidence for a
$\pi^0$ decay feature in the diffuse galactic $\gamma$-ray background,
even if the spectrum is harder than would be expected if CRs
throughout the Galaxy have the same spectrum as those observed
locally.\cite{Moskalenko:2004vh} There is also some evidence from Cas
A that the TeV emission originates from hadronic
interactions,\cite{Aharonian:2001mz} though at a much lower level than
would be expected if SNRs accelerated CRs with $\sim 10$\,\%
efficiency.

These clues suggest that the solution to the CR problem below the knee
of the CR spectrum at $\approx 3$ PeV and, indeed, the problem of the
origin of CRs at all energies, might be a consequence of the diversity
in the types of supernovae (SNe).  SNe are hardly uniform: the
simplest separation is between the explosively burning white dwarfs
(Type Ia), and the core collapse SNe (Type II and Types Ib and
Ic). Moreover, the ejecta speeds vary greatly,\cite{Weiler:2000rn}
from a few thousand km per second in Type II SNe with massive H
envelopes, to speeds of $\approx 10^5$ km/s in Type Ib/c SNe where the
H and He envelopes have been lost either to stellar winds or to Roche
lobe overflow to a binary companion.

The discovery that the GRBs which are observed today took place at
cosmological distances led to the development of the relativistic
fireball/blast-wave model\cite{Zhang:2003uk} that generalizes the
theory of SNe to stellar explosions with relativistic ejecta. The
coasting Lorentz factors of GRB outflows reach values of $10^2$ --
$10^3$. Associated with this discovery are other far-reaching
observational results\cite{pkw00} that tie GRBs to a subset of SNe:
GRBs are associated with massive stars in star-forming galaxies;
optically dark GRBs could be due to extreme reddening from large
quantities of dust and gas, as found in molecular cloud complexes
where massive stars are born; and delayed reddened excesses in the
late-time optical afterglow light curves appear to be SN emissions,
and can often be fit with a template light curve of the Type Ib/c SN
1998bw.

A final important discovery is that GRB outflows are in the form of
highly collimated jets. The evidence for beaming is inferred from
beaming breaks in optical afterglow light curves that occur when the
Doppler cone of the decelerating relativistic outflow is about the
same size as the jet cone. From the beaming breaks, one can infer that
the mean solid angle subtended by a GRB jet is about 1/500th of the
full sky.\cite{fra01} As a consequence, there are many hundreds of
GRBs events that take place for every one that is observed. By
performing the statistics of GRB sources, one finds that the rate of
GRBs in the Milky Way reaches $\approx 10$\% of the rate of Type 1b/c
SNe. All these lines of evidence suggest that GRBs are a species of
SNe.

\subsection{CRs from GRBs}

What does this mean for a complete model of cosmic rays? It is not
sufficient to provide an explanation for CRs with energies $\gsim
10^{18}$ eV without at least speculating on the origin of CRs with
energies from the knee to the ankle. Many explanations have been
suggested, including CR production from pulsars or extremely energetic
SNe in the Galaxy, acceleration at the Galactic wind termination
shock, or extragalactic models where CRs above the knee diffuse into
the Galaxy.\cite{Hoerandel:2004gv} A complete explanation based on
acceleration in SNe shocks would tie our understanding of CR
acceleration by SNe at GeV -- TeV energies with an origin of UHECRs in
the relativistic shocks associated with GRB SNe.

In the model of Ref.~[\cite{Wick:2003ex}], high-energy cosmic rays
(HECRs), defined as those with energies $\gsim 10^{14}$ eV, are
assumed to be accelerated at the shocks produced by the Type Ib/c SNe
that collapse to form GRBs.  From the evidence for beaming and the
association of GRBs with star-forming galaxies like the Milky Way, GRB
events are estimated to occur once every 3000 -- 10000 yrs in the
Galaxy. Relativistic shock acceleration in GRB blast waves are assumed
to inject power-law spectra of ions from a minimum energy $E_{\rm
min}\approx 10^{14}$~eV to a high-energy cutoff $E_{\rm max}\gsim
10^{20}$~eV. HECRs injected in the Milky Way diffuse and escape from
our Galaxy, and UHECRs with energies $\gsim 10^{17}$ -- $10^{18}$~eV
that have Larmor radii comparable to the size scale of the halo escape
directly from the Milky Way and propagate almost rectilinearly through
extragalactic space.  By the same token, UHECRs produced from other
galaxies can enter the Milky Way to be detected.  UHECRs formed in
GRBs throughout the universe travel over cosmological distances and
have their spectrum modified by energy losses, so an observer in the
Milky Way will measure a superposition of UHECRs from extragalactic
GRBs and HECRs produced in our Galaxy.

The model of Ref.~[\cite{Wick:2003ex}] fits the measured
KASCADE\cite{kam01} spectra of HECRs in the knee region, and the HiRes
data\cite{Abu-Zayyad:2002sf} at ultra-high energies. The fits imply
the HECR injection spectral index and luminosity function of GRBs.
The results show that GRBs must be strongly baryon-loaded, with the
testable prediction that GRBs will produce a detectable number of high
energy neutrino showers in a km-scale neutrino detector such as
IceCube.

\subsection{\label{GRBmodel}GRB Model for HE and UHECRs}

High energy CRs from a GRB in the Galaxy are assumed to propagate
diffusively as a result of pitch-angle scattering with
magneto-hydrodynamical (MHD) turbulence superposed on the Galactic
magnetic field.\cite{Wick:2003ex,Roulet:2003rr} The Larmor radius
$r_{\rm L}$ of a CR propagating in a magnetic field of strength
$B_{\mu{\rm G}} \equiv 1 \mu$G is $ \cong A \gamma_6/(Z B_{\mu{\rm
G}})$~pc, where $\gamma= 10^{6}\,\gamma_6$ is the Lorentz factor of a
CR with atomic mass $A$ and charge $Ze$.  The mean-free-path $\lambda$
between pitch-angle scatterings of a CR with Larmor radius $r_L$ is
assumed to be inversely proportional to the energy density in the MHD
spectrum at wave-number $k\sim r_L^{-1}$.  A two-component turbulence
spectrum is assumed with wave-number index $q = 3/2$ for a
Kraichnan-type spectrum at small wave number, and index $q = 5/3$ for
a Kolmogorov-type spectrum at large wave numbers.  The two components
give a $Z$-dependent break in the scattering mean-free-path $\lambda$
at energy $E_Z ({\rm PeV})\cong Z B_{\mu{\rm G}}b_{\rm pc},$ where we
take $B_{\mu{\rm G}}=3$, and find that $b_{\rm pc}=1.6$ is the
wavelength in parsecs of the MHD waves where the spectrum changes from
Kraichnan to Kolmogorov turbulence. The injection of turbulence at
large wave numbers is probably due to SN explosions in the Galaxy; the
small wave number turbulence could be a consequence of halo/disk
interactions (e.g., high-velocity clouds passing through the Galaxy's
disk).

The diffusion radius $r_{\rm dif} \cong 2\sqrt{\lambda c t/3}.$ When
$r\ll r_{\rm dif},$ the HECR differential number density $ n(\gamma; r,t)
\propto t^{-3/2}\times\gamma^{-p-3(2-q)/2}$.  The measured spectrum is
steepened by ${\frac{3}{2}}(2-q)$ units because the diffusion
coefficient $D\propto\lambda \propto \gamma^{2-q}$ for an impulsive
source.\cite{aav95} An injection spectrum with $p = 2.2$, as expected
from relativistic shock acceleration, gives a measured spectrum
$n_{Z,A}(\gamma; r,t)\propto
\gamma^{-s}$, with $s = 2.7$ at $E\ll E_Z$ and $s = 2.95$ at $E \gg
E_Z$. These indices are similar to the measured CR indices below and
above the knee energy. 

%%%%%%%%%%%%%%%%%
\begin{figure}[t]
\vskip-2.8in
\centerline{\epsfig{file=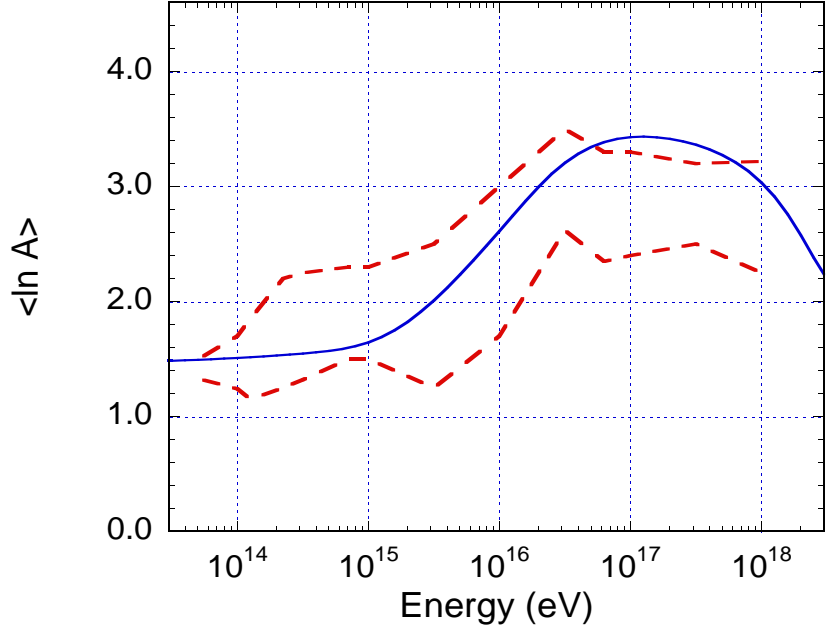,width=13.0cm,height=16.0cm }}
\caption[...]{ Solid curve shows the value of 
$\langle \ln A \rangle$ calculated from the model of Wick, Dermer, and
Atoyan (2004), where a single GRB in the Galaxy releases $10^{52}$
ergs in the form of a power-law spectrum of CRs with energies between
$\approx 10^{14}$ eV and $10^{20}$ eV.  In the model, a Galactic GRB
occurred $2.1\times 10^5$ years ago at a distance of 500 pc from
Earth.  The CRs isotropically diffuse via pitch-angle scattering with
an energy-dependent mean-free-path $\lambda$ determined by the MHD
turbulence spectrum. The dashed curves outline the expected range of
uncertainty in $\langle \ln A \rangle $ from the experimental data and
our knowledge of particle physics\cite{Hoerandel:2004gv}. The effect on
$\langle \ln A \rangle $ of extragalactic CRs at energies $\gsim
10^{17}$ is not included.
\label{fig:KASCADE}}
\end{figure}
%%%%%%%%%%%%%%%%%

\begin{figure}[t]
\vskip-2.2in
\centerline{\epsfig{file=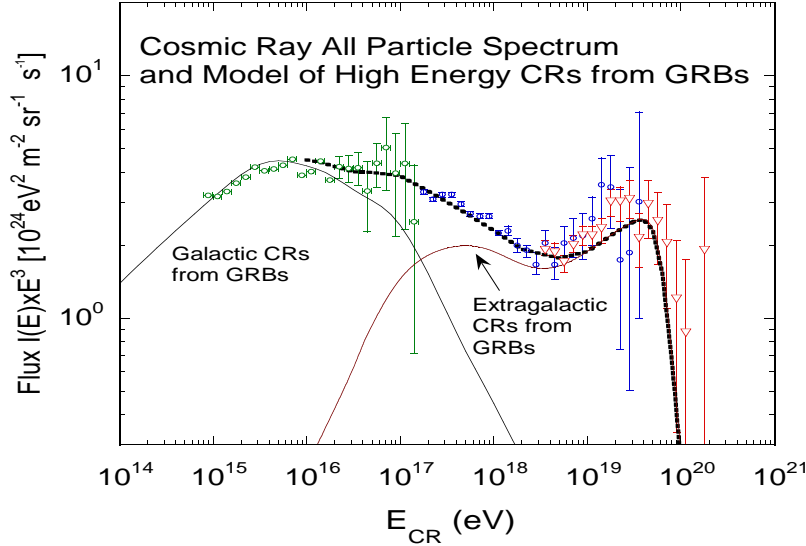,width=13.0cm,height=13.0cm}}
\caption[...]{Fit of the model of Ref.\ [\cite{Wick:2003ex}]
to the KASCADE ($< 2\times 10^{17}$ eV), HiRes-II Monocular (circles),
and HiRes-I Monocular (inverted triangles) data, assuming that
GRBs inject a power law distribution of particles with index $p = 2.2$
from $10^{14}$ eV to a maximum energy with an exponential cutoff at
$E_{\rm max}=10^{20}$~eV.  A minimum $\chi^2$ routine was used to find the
best fit model which has a $\chi^2_r = 1.03$. This fit implies that
the transition from Galactic to extragalactic CRs occurs near the
second knee at $10^{17.6}$~eV, and that the ankle ($10^{18.5}$~eV) is
a feature associated with photo-pair production. The heavy dark points
shows the best fit model, and the galactic and extragalactic
components are shown separately.
\label{fig:cr4}}
\end{figure}

Figure~\ref{fig:KASCADE} shows the predicted mean logarithmic mass
$\langle \ln A \rangle $ for CRs which diffuse from a GRB that
occurred $2.1\times 10^5$ years ago at a distance of 500 pc from
Earth.  For comparison, the range of $\langle \ln A \rangle $ derived
from direct measurements using a phenomenological model of particle
interactions is shown.\cite{Hoerandel:2004gv} As can be seen, the
model gives a good fit to the data at energies $\ll 10^{17}$ eV, which
is expected because the model also provides good fits to the
individual spectra of CR ions measured with KASCADE. The model
displays an excess abundance of heavy ions at higher energies, but the
calculated value of $\langle \ln A \rangle$ shown here does not
include contributions from extragalactic CRs, which become important
above $\gsim 10^{17}$ eV (see Fig.~\ref{fig:cr4}).  If the
extragalactic component is primarily CR protons, then this discrepancy
is resolved.

Good fits to knee data were obtained with a break in the turbulence
spectrum at wave numbers corresponding to $\approx 1$ pc. Composition
enhancements by a factor of 50 and 20 for C and Fe, respectively, over
Solar photospheric abundances are also required.  The likelihood for
such an event is reasonable, and the corresponding anisotropy of the
CRs from a single source is shown to be consistent with
observations.\cite{Wick:2003ex}

UHECRs produced by GRBs throughout the universe lose energy
adiabatically during the expansion of the universe, and through
photo-pair and photo-pion production on the CMB. The loss processes
produce features in the UHECR flux from distant ($z\gsim 1$) GRB
sources at characteristic energies $\sim 4\times 10^{18}$~eV and $\sim
5\times 10^{19}$~eV from photo-pair and photo-pion processes,
respectively.  Comparison of the integrated and evolved UHECR spectrum
with data depends on how the luminosity density of GRB sources evolve
through cosmic time.  In view of the evidence that GRBs are associated
with massive stars, the GRB luminosity density is assumed to be
proportional to the star formation rate (SFR) history of the universe
as traced by the blue and UV luminosity density of distant
galaxies. The lowest possible SFR is expected to follow the evolution
of the observed optical/UV luminosity density, and a higher rate to
follow this rate with allowance for extinction
corrections.\cite{Blain99}

Figure~\ref{fig:cr4} shows the best fit model to the all-particle
spectrum from below the knee of the CR spectrum to the highest
energies.  The best fit model to the data has an injection index $p =
2.2$, a high-energy exponential cutoff energy of $10^{20}$ eV, and a
SFR which follows the higher extinction-corrected rate.  The
transition between galactic and extragalactic CRs is found in the
vicinity of the second knee ($10^{17.6}$~eV), consistent with a
heavy-to-light composition change.\cite{Bird:yi} The ankle
($10^{18.5}$~eV) is interpreted as a suppression from photo-pair
losses, analogous to the GZK suppression at higher energies.

%%%%%%%%%%%%%%%%%
%%%%%%%%%%%%%%%%%

Fits to the data define the local luminosity density of GRBs for
producing HECRs. The GRB HECR luminosity density is found to be $\cong
70\times 10^{44} ~\rm{erg}~\rm{Mpc}^{-3}\rm{yr^{-1}}$, so that this
model implies that GRBs inject considerably more energy in the
form of nonthermal hadrons than electrons.  Thus GRB blast waves are
baryon-loaded by a factor $\approx 60$ -- 100 compared to the energy
radiated by GRBs in the form of hard X-ray and soft $\gamma$-ray
emission.  For the large baryon load required for this model,
calculations show that 100 TeV -- 100 PeV neutrinos could be detected
several times per year from all GRBs with kilometer-scale neutrino
detectors such as IceCube.\cite{Wick:2003ex,Dermer:2003zv}  Detection 
of even 1 or 2
neutrinos from GRBs with IceCube or a northern hemisphere neutrino
detector would unambiguously demonstrate the high nonthermal baryon
load in GRBs, and would provide compelling support for this scenario
for the origin of CRs.

Finally, we note that these calculations have implicitly assumed that
the Fly's Eye/HiRes measurements of the UHECR spectrum are correct.
If a post-GZK excess is confirmed, then physics beyond the standard
model could be required, as discussed in Sec.~\ref{Neutrino}.

\subsection{The Last GRB in the Galaxy and the CR Excess from the GC}

As discussed in Sec.~\ref{john}, the AGASA\cite{Hayashida:1998qb} and
the SUGAR\cite{Bellido:2000tr} experiments observed an excess of CRs
from the direction of the GC. The SUGAR data suggest a point source to
within their spatial resolution, while AGASA shows an extended
source. The excess starts to be significant around $10^{17.5}$ eV,
peaks near $10^{18}$ eV, and cuts off sharply at about $10^{18.5}$ eV.
The flux of the excess particles represents a luminosity of particles
beyond $10^{18}$ eV of about $4 \times 10^{30}$~erg/s.  The GC  anisotropy 
is then very suggestive of neutrons as candidate
primaries, because the directional signal requires relatively-stable
neutral primaries, and time-dilated neutrons can reach the Earth from
typical Galactic distances when the neutron energy exceeds
$10^{18}$~eV.

A novel hypothesis to explain the GC excess involving GRBs was
presented in this Parallel Session by Gustavo Medina
Tanco.\cite{M-Tanco} In order to estimate the remaining traces of any
CR activity produced by the last GRB in the galaxy, one has to take
into account several considerations:

\noindent{\it (i)} The UHECRs escaping the GRB fireball
\begin{equation}
N_0(E>10^{18}~{\rm eV}) \sim 10^{-2}\,\, N_0 
(E> E_{\rm min}) 
\end{equation}
are mostly neutrons, because protons are captive in the magnetic field
and suffer extensive adiabatic losses on the way
out.\cite{Rachen:1998fd}\footnote{We remind the reader that the
differential injection spectrum of GRBs $\propto E^{-2.2}.$} Some of
these neutrons will decay into protons within the GC thin disk-like
($r \sim 3$~kpc) region of high interstellar medium density and high
star formation rate. The population of secondary protons would then be
captured by the strong $B$-field near the GC, attaining diffusion with
a residence time scale of about $T \sim 10^{5}$~yr.  At the end of
this time, about 1/300 protons are able to avoid leakeage.  The
trapped protons,
\begin{equation}
N_T (E> 10^{18}~{\rm eV}) = N_0 (E>10^{18}~{\rm eV}) \,\, 
\left(1-e^{-\frac{r\,m_n}{E\,\overline\tau}}\right) / 300\,\,,
\end{equation} 
can then be turned back into neutrons by interaction with nuclei in
the interstellar medium with probability of $5 \times 10^{-2}.$ Here,
$m_n$ and $\overline\tau$ are the neutron mass and lifetime,
respectively.

\noindent{\it (ii)} The formation of the 
$n \rightarrow p$ reservoir depends on the GRB rate in the inner
Galaxy times the probability that a GRB jet points more o less along
the direction of the GP. The latter is estimated to be about 50\,\%.

\noindent{\it(iii)} The total CR production by a single GRB 
is $\approx 10^{51}$~erg.\cite{Pugliese:1999df}
 
Putting all this together, the observed anisotropy in the direction of
the GC can be easily fitted by the neutrons produced in the GRB
reservoir, that ultimately travel unscathed to
Earth.\cite{Biermann:2004hi} Arguably, if the anisotropy messengers
are neutrons, then those with energies below $10^{18}$~eV will decay
in flight, providing a flux of cosmic antineutrinos above 1 TeV that
should be observable at kilometer-scale neutrino
telescopes.\cite{Anchordoqui:2003vc}

\subsection{UHECRs, EMBHs, and GRBs}

An alternative explanation for the GRB phenomenon has been recently
put forward.\cite{Preparata:1998rz} In this model the GRB explosion is
due to the rapid discharge of an electromagnetic black hole
(EMBH). This chain reaction occurs when a massive star collapses into
an EMBH able to produce a dyadosphere, i.e., the EMBH reaches the
critical field for the Schwinger process, ${\mathcal E_c} = m_e^2 c^3/
\hbar e \simeq 1.3 \times 10^{18}$~V/m.\cite{Damour:1974qv} The $e^\pm$ 
pairs created through vacuum polarization then promptly annihilate
into the GRB. This model explains the time variability, the spectra,
and the GRB afterglows to a very high level of accuracy.\cite{Ruffini}
Based on the EMBH-GRB hypothesis, in this Parallel Session Alvise
Mattei\cite{Mattei} presented a mechanism to accelerate the ionized
hydrogen atoms surrounding the death star to ultra high energies.

\begin{figure} [t]
\postscript{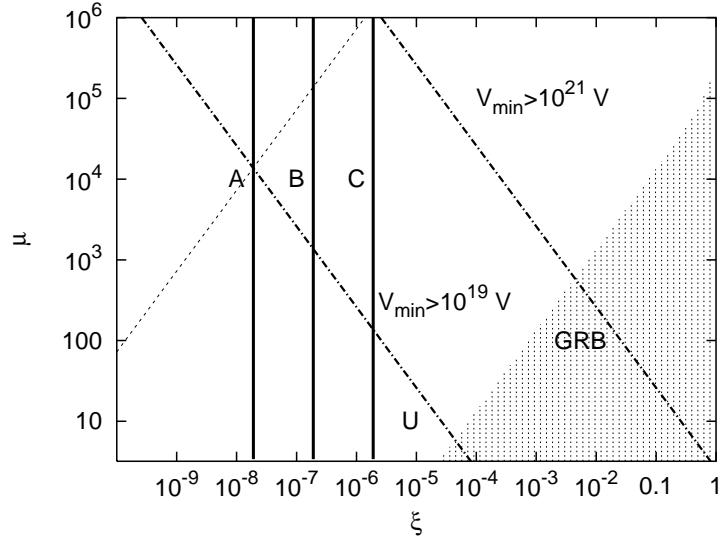}{0.80}
\caption{Limits on charge and mass ratio for UHECR emission. EMBHs 
lying in the shaded zone develop a dyadosphere. EMBHs above the
dotted line cannot activate the ionization mechanism. EMBHs on the
right hand side of the dashed-dotted lines produce particles at lower
energies between $10^{19}$~eV and $10^{21}$~eV. The vertical solid lines
indicate the maximum energy limit for proton acceleration, A, B, and C
corresponds to $E_{\rm max} > 10^{19}$~eV, $E_{\rm max} > 10^{20}$~eV,
and $E_{\rm max} > 10^{21}$~eV; respectively.  (This figure is
courtesy of Alvise Mattei).
\label{embh}}
\end{figure}

In this set-up protons would be accelerated outward in a very short
time along a straight line by the induced electric field. Larmor
losses would overcome gains if the energy increases as $dE/dx \sim
10^{21}$ MeV/m. Thus, protons do not suffer catastrophic energy losses
if the EMBH cannot produce a dyadosphere.  In terms of the mass
$M_{\rm BH}$ and the charge $Q_{\rm BH}$ of the BH, the condition for
dyadosphere formation near the BH horizon (of radius $r_{\rm H}$),
${\mathcal E} (r_{\rm H}) < {\mathcal E_c}$, is found to be
\begin{equation} \mu <  6 \times 10^5 \, \xi\,\, 
\left(1 + \sqrt{1 -\xi^2}\right)^{-2}
\label{Schw}
\end{equation}
where $\mu \equiv M_{\rm BH}/M_\odot$ and $\xi \equiv Q_{\rm
BH}/Q_{\rm max}$. In order to produce a baryon reservoir, the
electromagnetic field has to be able to ionize hydrogen atoms, i.e., a
ionization potential of 13.6~V must be active in a distance of about a
Bohr radius. This condition can be re-written as ${\mathcal E} (r_{\rm
H}) \gsim 2.5 \times 10^{11}$~V/m, or equivalently
\begin{equation}
\mu \leq 2.9 \times 10^{12} \,\xi\,\, \left(1 + \sqrt{1 -\xi^2}\right)^{-2} \,.
\end{equation}
The limits impossed by these constraints are summarized in Fig.~\ref{embh}. 

In summary, an EMBH would lose its charge by emitting UHECRs if the BH
does not reach the the critical charge-to-mass ratio given in
Eq.~(\ref{Schw}). This type of BH has been classified as pure or
U-EMBH. The BH population that has recently exploded as a GRB ends up
near the dyadosphere zone with residual charge, and so UHECR emission
is still possible until complete discharge.

%%%%%%%%%%%%%%%%%%%%%%
\begin{figure}[t] 
\vspace{-4ex}
\centerline{\epsfxsize=3.6in\epsfbox{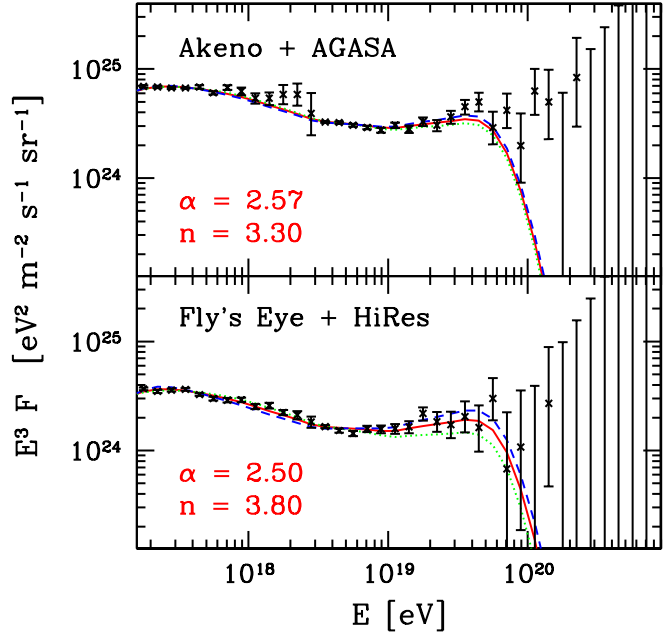}}   
\caption[...]{
Ultrahigh energy cosmic ray data with their statistical errors (top:
combination of Akeno\cite{Nagano:1991jz} and AGASA\cite{Takeda:1998ps}
data; bottom: combination of Fly's Eye\cite{Bird:yi} and
HiRes\cite{Abu-Zayyad:2002ta} data) and the predictions arising from a
power-law emissivity distribution~(\ref{source-emissivity})
corresponding to sources which are uniformly distributed at
cosmological distances.  The best fits between $E_-=10^{17.2}$~eV and
$E_+=10^{20}$~eV are given by the solid lines and correspond to the
indicated values of the parameters $\alpha$ and $n$ in the source
emissivity distribution.  The 2-sigma variations corresponding to the
minimal (dotted) and maximal (dashed) fluxes are also shown.  Other
parameters of the analysis were $E_{\rm max} = 3 \times 10^{21}$~eV,
$z_{\rm min} = 0.012$, and $z_{\rm max}=2$.  From
Ref.~[\cite{Fodor:2003ph}].
\label{uhecr-data}}
\end{figure}
%%%%%%%%%%%%%%%%%%%%% 

\section{\label{Neutrino}Strongly Interacting Neutrinos}

We have noted in the Introduction that if the highest energy cosmic
rays are nucleons (or nuclei), if their sources are indeed uniformly
distributed at cosmological distances, and if their injection spectra
are power-laws in energy -- a reasonable assumption, in view of the
measured spectrum in Fig.~\ref{fig:cr4} which appears to be
approximately of (broken) power-law type over many order of magnitude
in energy -- then their total flux arriving at Earth should show a
pronounced drop above the GZK cutoff $E_{\rm GZK} \approx 4\times
10^{19}$~eV ($1 \times 10^{20}$~eV, for nuclei).  This is due to the fact that,
above this energy, the universe becomes opaque to high energy nucleons
(and nuclei), due to inelastic hadronic scattering processes with the
CMB photons. The GZK cutoff is, however, not seen in the data, at
least not in a significant manner
(cf. Fig.~\ref{uhecr-data}). Correspondingly, the events above
$10^{20}$~eV in Fig.~\ref{uhecr-data} should originate from small
distances below $50$~Mpc, the typical interaction length of nucleons
above $E_{\rm GZK}$. However, no source within a distance of $50$~Mpc
is known in the arrival directions of the post-GZK events.  The basic
puzzle is: if there are no large intervening magnetic fields and the
sources of ultrahigh energy cosmic rays are indeed at cosmological
distances, how could they reach us with energies above $10^{20}$~eV?

At the relevant energies, among the known particles only neutrinos can
propagate without significant energy loss from cosmological distances
to us.  This fact leads naturally to scenarios invoking hypothetical
-- beyond the Standard Model of elementary physics -- strong
interactions of ultrahigh energy cosmic neutrinos,\cite{Beresinsky:qj}
whose modern incarnation we review in this section.\cite{Fodor:2004tr}

Such scenarios are based on the observation that the flux of neutrinos
originating from the decay of the pions produced during the
propagation of nucleons through the
CMB\cite{Beresinsky:qj,Stecker:1979ah} -- the cosmogenic neutrinos --
shows a nice agreement with the observed UHECR flux above $E_{{\rm
GZK}}$.  Assuming a large enough neutrino-nucleon cross-section at
these high energies, these neutrinos could initiate extensive air
showers high up in the atmosphere, like hadrons, and explain the
existence of the post-GZK events.\cite{Beresinsky:qj} This large
cross-section is usually ensured by new types of TeV-scale
interactions beyond the Standard Model, such as arising through
gluonic bound state leptons,\cite{Bordes:1997bt} through TeV-scale
grand unification with leptoquarks,\cite{Domokos:2000dp} through
Kaluza-Klein modes from compactified extra
dimensions,\cite{Domokos:1998ry} (see, however,
Ref.~[\cite{Kachelriess:2000cb}]), or through $p$-brane production in
models with warped extra dimensions\cite{Ahn:2002mj} (see, however,
Ref.~[\cite{Anchordoqui:2002it}]); for earlier and further proposals,
see Ref.~[\cite{Domokos:1986qy}] and Ref.~[\cite{Barshay:2001eq}],
respectively.

In Refs.~[\cite{Fodor:2003bn,Fodor:2004tr}], a detailed statistical
analysis of the agreement between observations and predictions from
such scenarios was presented.  Moreover, an example was emphasized
which -- in contrast to previous proposals -- is based entirely on the
Standard Model of particle physics.  It exploits non-perturbative
electroweak instanton-induced processes for the interaction of
cosmogenic neutrinos with nucleons in the atmosphere, which may have a
sizeable cross-section above a threshold energy $E_{\rm th}={\mathcal
O}( (4\pi m_W/\alpha_W )^2)/(2 m_p) = {\mathcal O}( 10^{18})$~eV,
where $m_W$ denotes the W-boson mass and $\alpha_W$ the electroweak
fine structure
constant.\cite{Aoyama:1986ej,Morris:1991bb,Ringwald:2002sw}

The scenario is based on the assumption of a power-law emissivity
distribution corresponding to uniformly distributed sources and thus
quite consistent with the GRB model of CR origin from
Sec.~\ref{GRBmodel}.  The emissivity is defined as the number of
protons per co-moving volume per unit of time and per unit of energy,
injected into the CMB with energy $E_i$ and characterized by a
spectral index $\alpha$ and a redshift ($z$) evolution index $n$,
\begin{equation}
\label{source-emissivity}
{\mathcal L}_p =j_0\,E^{-\alpha}_i\,
\left(1+z\right)^n\,\theta(E_{\rm max}-E_i)\,
\theta(z-z_{\rm min})\,\theta(z_{\rm max}-z)\,.
\end{equation}
Here, $j_0$ is a normalization factor, which will be fixed by the
observed flux.  The parameters $E_{\rm max}$ and $z_{\rm min/max}$
have been introduced to take into account certain possibilities such
as the existence of a maximal energy, which can be reached through
astrophysical accelerating processes in a bottom-up scenario, and the
absence of nearby/very early sources, respectively. They turn out to
be quite insensitive to the specific choice for $E_{\rm max}$, $z_{\rm
min}$, and $z_{\rm max}$, within their anticipated values. The main
sensitivity arises from the spectral parameters $\alpha$ and $n$, for
which 1\, and 2\,$\sigma$ confidence regions have been
determined. Note, that $n\sim 3, z_{\rm max}\sim 2$ mimicks the SFR
history of the universe mentioned also in connection with GRBs in
Sec.~\ref{GRBmodel}.

After propagation through the CMB, the protons from
Eq.~(\ref{source-emissivity}) will have energies below $E_{{\rm
GZK}}$, so they can well describe the low energy part of the UHECR
spectrum.  The cosmogenic neutrinos interact with the atmosphere and
thus give a second component to the UHECR flux, which describes the
high energy part of the spectrum. The relative normalization of the
proton and neutrino fluxes is fixed in this scenario, so the low and
high energy parts of the spectrum are explained simultaneously without
any extra normalization.  Details of this analysis can be found in
Ref.~[\cite{Fodor:2003bn}].

%%%%%%%%%%%%%%%%%%%%%%
\begin{figure}[t] 
\vspace{-0.4ex}
\centerline{\epsfxsize=3.6in\epsfbox{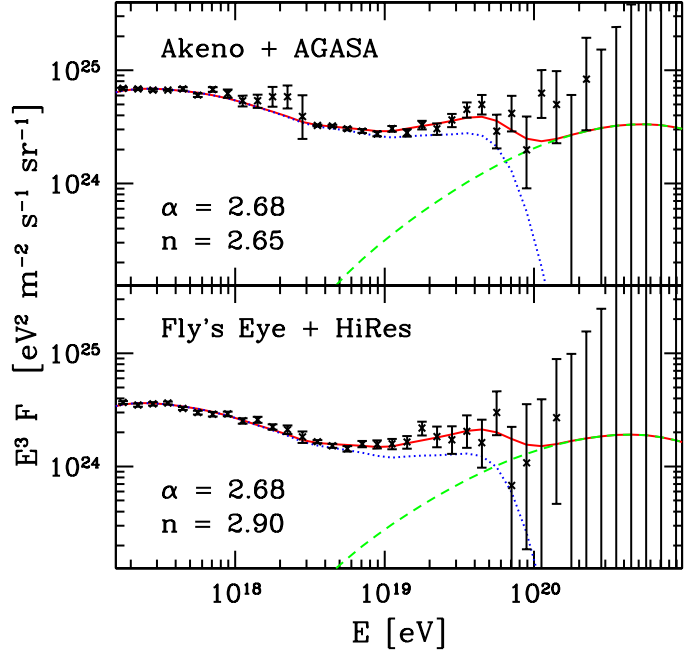}}
\caption[...]{
Ultrahigh energy cosmic ray data (Akeno + AGASA on the upper panel and
Fly's Eye + HiRes on the lower panel) and their best fits (solid)
within the electroweak instanton scenario, for $E_{\rm max}=3\times
10^{22}$~eV, $z_{\rm min}=0.012$, $z_{\rm max}=2$, consisting of a
proton component (dotted) plus a cosmogenic neutrino-initiated
component (dashed).  From Ref.~[\cite{Fodor:2003bn}].
\label{fit}}
\end{figure}
%%%%%%%%%%%%%%%%%%%%%

Figure~\ref{fit} shows the best fits for the AGASA and for the HiRes
UHECR data. The best fit values are $\alpha=2.68(2.68)$ and
$n=2.65(2.9)$, for AGASA (HiRes), within the electroweak instanton
scenario.  One can see very nice agreement with the data within an
energy range of nearly four orders of magnitude.  The fits are
insensitive to the value of $E_{\rm max}$ as far as one chooses a
value above $\approx 3\times 10^{21}$~eV. The maximum injection
energy, however, is constrained by EGRET measurements of the diffuse
gamma ray flux.\cite{Semikoz:2003wv} This is because the photons
produced via $\pi^0$ decay are degraded in energy due to interactions
with the universal radition backgrounds. It is noteworthy that the use
of a 2-component model allows $E_{\rm max} \approx 3 \times
10^{22}$~eV without violating the new EGRET
bounds.\cite{Strong:2003ex}

The shape of the curve between $10^{17}$~eV and $10^{19}$~eV is mainly
determined by the redshift evolution index $n$. At these energies the
universe is already transparent for protons created at $z\approx 0$,
while protons from sources with larger redshift accumulate in this
region.  The more particles are created at large distances -- i.e. the
larger $n$ is -- the stronger this accumulation should be. In this
context, on may note that the data seem to confirm the implicit
assumption that the extragalactic uniform UHECR component begins to
dominate over the galactic one already at $\approx 10^{17}$~eV. If
one, alternatively, starts the fit only at $10^{18.5}$~eV --
corresponding to the assumption that the galactic component dominates
up to this energy -- one finds, however, also a very good fit, with a
very mild dependence on $n$ and the same best fit values for $\alpha$,
with a bit larger uncertainties.
The peak around $4\times 10^{19}$~eV in Fig.~\ref{fit} shows the
accumulation of particles due to the GZK effect. Neutrinos start to
dominate over protons at around $10^{20}$~eV.

%%%%%%%%%%%%%%%%%%%%%%
\begin{figure}[t] 
\centerline{\epsfxsize=2.8in\epsfbox{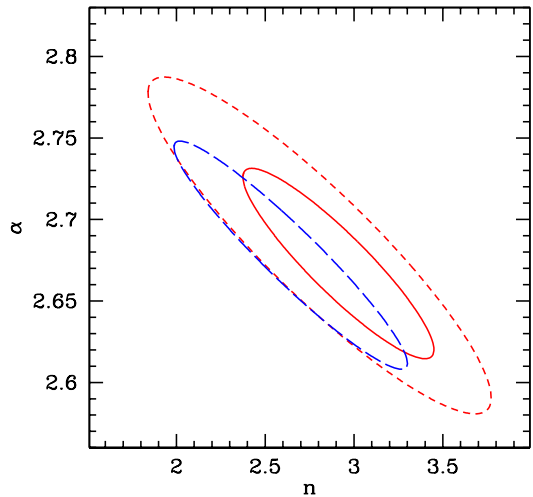}}   
\vspace{-0.5cm}
\caption[...]{
Confidence regions in the $\alpha$--$n$ plane for fits to the Akeno +
AGASA data (2-sigma (long dashed)) and to the Fly's Eye + HiRes data
(1-sigma (solid); 2-sigma (short-dashed)), respectively, within the
electroweak instanton scenario, for $E_{\rm max}\gwig 3\times
10^{21}$~eV, $z_{\rm min}\geq 0$, $z_{\rm max}=2$.  From
Ref.~[\cite{Fodor:2003bn}].
\label{confidence}}
\end{figure}
%%%%%%%%%%%%%%%%%%%%%

Figure~\ref{confidence} displays the confidence regions in the
($\alpha$,$n$) plane for AGASA and HiRes. The scenario is consistent
on the 2-sigma level with both experiments. For HiRes, the
compatibility is even true on the 1-sigma level.  It is important to
note that both experiments favor the same values for $\alpha$ and $n$,
demonstrating their mutual compatibility on the 2-sigma level (see
also Ref.~[\cite{DeMarco:2003ig}]).  If one ignores the energy
uncertainty in the determination of the goodness of the fit, they turn
out to be inconsistent.

Finally, let us emphasize that the same fit results are valid for all
strongly interacting neutrino scenarios, as long as the
neutrino-nucleon cross-section has a threshold-like behavior as in the
case of electroweak instanton-induced processes, with a neutrino
threshold energy $\lwig\, 4\times 10^{19}$~eV and a cross-section
$\gwig\, 1$~mb above threshold. It is also important to note that the
energy requirements on the sources of the primary protons are
comparatively mild. To obtain a good fit, one needs $E_{\rm
max}\gwig\, 3\times 10^{21}$~eV. There are several astrophysical
source candidates with CR-emission up to this
energy;\cite{Anchordoqui:2002hs} notably, as we discussed in Sec.~3,
GRBs can provide the necessary conditions to accelerate protons to the
required energies by conventional shock acceleration. Moreover, they
naturally provide the power-law source emissivity distribution assumed
in Eq.~(\ref{source-emissivity}).

The predicted ultrahigh energy cosmic neutrino component can be
experimentally tested at cosmic ray facilities by studying the zenith
angle dependence\cite{Morris:1991bb,Berezinsky:kz} of the events in
the range $10^{18}$ -- $10^{20}$~eV and by analyzing possible
correlations with distant astrophysical sources. Additionally, one can
look for bumps in neutrino-initiated shower spectra at neutrino
telescopes such as IceCube.\cite{Han:2003ru} As laboratory tests, one
may search for enhancements in (quasi-)elastic lepton-nucleon
scattering\cite{Goldberg:1998pv} or for signatures of QCD
instanton-induced processes in deep-inelastic
scattering,\cite{Balitsky:1993jd} e.g. at HERA.\cite{Adloff:2002ph}

In summary, strongly interacting neutrino scenarios provide a viable
and attractive solution to the UHECR puzzle and may be subject to
various crucial tests in the foreseeable future.

\section{Concluding Remarks}

We have reviewed some of the interesting ideas currently invoked to
explain the origin of UHECRs. We expect future CR experiments, such as
the Pierre Auger Observatory,\cite{Abraham} would clarify the
confusing experimental situation and shed light on validity of models
discussed in this report.

The Pierre Auger Observatory is designed to measure the energy and
arrival direction of UHECRs with unprecedented precision. It will
consist of two sites, one in the Northern hemisphere and one in the
Southern, each covering an area $S \approx 3000$~km$^2$.  The Southern
site is currently under construction while the Northern site is
pending. Once complete, these two sites together will provide the full
sky coverage and well matched exposures which are crucial for
anisotropy analyses. The base-line design of the detector includes a
ground array consisting of 1600 water \v{C}erenkov detectors
overlooked by 4 fluorescence eyes. The angular and energy resolutions
of the ground arrays are typically less than $1.5^\circ$ and less than
20\%, respectively. The detectors are designed to be fully efficient
out to $\theta_{\rm max} = 60^\circ$ beyond $10^{19}$~eV.  In 10~yr of
running the two arrays will collect $\approx 4000$ events above
$10^{19.6}$~eV. Such statistics will enable us to solve the GZK
puzzle.

\section*{Acknowledgments}
We would like to thank Mario Novello, Santiago Perez Bergliaffa and
Remo Ruffini for this very fruitful conference. The work of LAA has
been partially supported by the US National Science Foundation (NSF)
under grant No.  PHY-0140407. The work of CDD is supported by the
Office of Naval Research and {\it GLAST} Science Investigation Grant
No.\ DPR-S-1563-Y.

%%%%%%%%%%%%%%%%%%%%%%%%%%%%%%%%%%%%%%%%%%%%%%%%%%%%%%

\end{document}